\documentclass{elsart41}
\def\ket#1{\left\vert #1 \right\rangle}
\def\bra#1{\left\langle #1 \right\vert}
\newcommand{\be}{\begin{equation}}
\newcommand{\ee}{\end{equation}}
\newcommand{\bea}{\begin{eqnarray}}
\newcommand{\eea}{\end{eqnarray}}

\usepackage{graphics}
\usepackage{graphicx}
\usepackage{amssymb}
\begin{document}

\begin{frontmatter}

\vspace{-4cm}
\title{Finite-curvature scaling in optical lattice systems}

\author[AA]{C. Hooley},
\ead{c.hooley@bham.ac.uk}
\author[BB]{J. Quintanilla\corauthref{Name2}},\,
\ead{j.quintanilla@rl.ac.uk}

\address[AA]{School of Physics and Astronomy, University of Birmingham,  
Edgbaston, Birmingham B15 2TT, U.K.}  
\address[BB]{ISIS facility, Rutherford Appleton Laboratory, Chilton, Didcot,
Oxfordshire OX11 0QX, U.K.}

\corauth[Name2]{Corresponding author. Tel: +44 (0)123 544 6353
Fax: 5720}

\begin{abstract}

We address the problem posed by the inhomogeneous trapping fields
when using ultracold fermions to simulate strongly correlated
electrons.  As a starting point, we calculate the density of states
for a single atom. Using semiclassical arguments, we show that this
can be made to evolve smoothly towards the desired limit by varying
the curvature of the field profile. Implications for mutually
interacting atoms in such potentials are briefly discussed.

\end{abstract}

\begin{keyword}
Semiclassical theories and applications \sep
Degenerate Fermi gases  \sep
Lattice fermion models (Hubbard model, etc.)
\PACS    
03.65.Sq ;
03.75.Ss ;	
71.10.Fd 	
\end{keyword}

\end{frontmatter}


Since the first observation of Bose-Einstein condensation in a trapped gas of
ultra-cold atoms \cite{Anderson:1995},  there has been an
explosion of experimental and theoretical work on such systems.  More recent
developments include adding to the trapping fields a
laser standing wave (a so-called `optical lattice').  This generates an
effective potential for the atoms which is given by
\be
V(x) = V_0 \cos \left( \frac{2\pi x}{a} \right) + V_{\rm trap}(x),
\label{potential}
\ee
where $a$ is the spatial period of the optical lattice, $V_0$ its strength,
and $V_{\rm trap}(x)$ the effective potential due to the trapping fields. 
(The generalisation of (\ref{potential}) to more than one dimension is
straightforward.) Recently, degenerate fermions have been loaded into such 
optical lattices \cite{Esslinger}, opening up the possibility of 
simulating strongly correlated
electrons. Here we explore this prospect by considering, as a starting point,
the single-atom density of states (DOS). 

\begin{figure}[!ht]
\begin{center}
\includegraphics[width=0.4\textwidth]{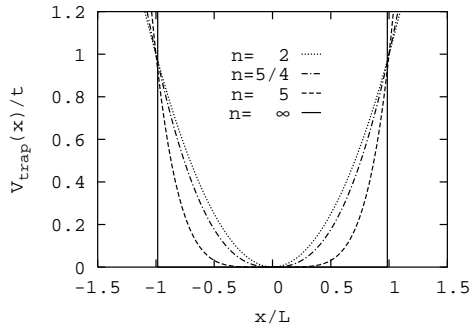}
\end{center}
\caption{Confining potential for various values of $n$, as indicated.}
\label{homdos1d}
\end{figure}

Perhaps the simplest case is that in which the optical lattice is very strong. 
In this case, we can replace the full Hamiltonian,
\be
H_{\rm full} = \frac{{\hat p}^2}{2m} + V_0 \cos \left( \frac{2\pi x}{a} \right)
+ V_{\rm trap}(x),
\ee
by a tight-binding version,
\begin{eqnarray}
H  &=& \sum_j \big\{ -t \ket{j} \bra{j+1} -t \ket{j+1} \bra{j}  \nonumber
\\ & &  \hspace{3cm}+ V_{\rm trap}(ja) \ket{j} \bra{j} \big\}, \label{tbham}
\end{eqnarray}
where $\ket{j}$ is a state in which the atom is in its local ground state on
site $j$ of the lattice.  

\begin{figure}[!ht]
\begin{center}
\includegraphics[width=0.4\textwidth]{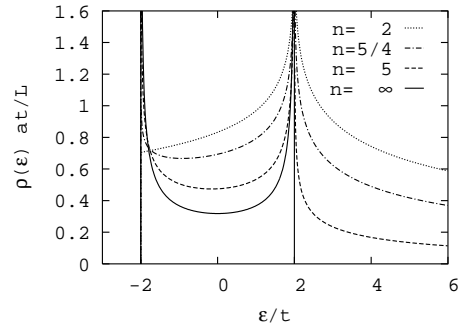}
\end{center}
\caption{The density of states for various values of $n$, as indicated.  As $n$
increases, the 
singularity at $\epsilon=-2t$ strengthens, while the 
one at $\epsilon=2t$ becomes increasingly asymmetric.}
\label{differentn}
\end{figure}

In models of electron systems, one usually assumes the confining 
potential to be a `box': $V_{\rm trap}(x)=0$ for $|x|<L$, $\infty$
otherwise (the solid line of Fig.\ \ref{homdos1d}). In the bulk
limit, 
\begin{equation}
L\to\infty,
\label{bulk}
\end{equation} 
$V_{\rm trap}(x)$ vanishes everywhere and 
one obtains the usual tight-binding DOS (the solid line of 
Fig.~\ref{differentn}).

On the other hand, ultracold atoms are usually subject to a harmonic
trap:  $V_{\rm trap}(x) = t (x/L)^2$. The bulk limit (\ref{bulk})
corresponds again to a vanishingly weak trapping potential, yet the
DOS is qualitatively different. It is given by 
the dotted line in Fig.\ \ref{differentn}, as can be shown either via 
a WKB approach \cite{Hooley-Quintanilla-04} or a  `local-density 
approximation' (LDA):
\be
\rho(\epsilon) = \sum_j \rho_{\rm hom}(\epsilon-V_{\rm trap}(ja)), \label{lda}
\ee
where $\rho_{\rm hom}(\epsilon)$ is the homogeneous DOS shown in
Fig.~\ref{homdos1d}.  (We prove elsewhere that these two methods are 
equivalent.) 

This state of affairs is unsatisfactory, in the sense that one would like to be
able to interpolate smoothly between the harmonic scenario realised in atom-traps and
the open boundary conditions of usual solid-state physics --- particularly if one hopes to
use the former to simulate the latter.  We propose to achieve this by
varying the form of the power law
at $x=0$.  Consider the family of potentials
\be
V_{\rm trap}(x) = t \left\vert \frac{x}{L} \right\vert^n. \label{ntrap}
\ee
These reduce to the two cases studied above for $n\to\infty$ and
$n=2$, respectively.  Let us therefore study the behaviour of the single-atom
DOS of (\ref{tbham}), with trapping potential (\ref{ntrap}), as a
function of $n$.  We call this progression through the family (\ref{ntrap})
`finite-curvature scaling', by analogy with the finite-size scaling procedure
used in numerical simulations \cite{Privman-90}.

We may obtain some information via straightforward asymptotics.  Firstly, in the
case $\epsilon \gg 2t$, we may neglect the first two terms in (\ref{tbham}). 
What remains is just the $V_{\rm trap}$ term, the eigenstates of which are
clearly the position eigenstates $\ket{j}$.  The
`dispersion relation' is then $\epsilon=t \vert x/L \vert^n$, implying that
$\vert x \vert = L (\epsilon/t)^{1/n}$, from which
the DOS may readily be obtained:
\be
\rho(\epsilon) \equiv \frac{dn}{d\epsilon} = \frac{dn/d \vert x
\vert}{d\epsilon/d \vert x \vert}
= \frac{2}{a} \frac{d \vert x \vert}{d\epsilon}
= \frac{2L}{atn} \left(\frac{\epsilon}{t}\right)^{(1/n)-1}.
\ee
This gives $\epsilon^{-1/2}$ behaviour in the harmonic case
\cite{Hooley-Quintanilla-04} in agreement with recent experiments
\cite{Ott-et-al-04}.  Note that it also
predicts $\epsilon^{-1}$ behaviour in the $n \to \infty$ limit; this prediction,
though correct, is irrelevant, because the spectral weight at $\epsilon > 2t$ goes to
zero as $n \to \infty$.

Secondly, we may look at the case $\epsilon \approx -2t$:\ these
are the states very near the bottom of the band.  For them, the tight-binding
form of the DOS can be expanded around $k=0$:
$\epsilon_{\rm hom}(k) = -2t \cos (ka) \approx -2t + t a^2 k^2$,
corresponding to the kinetic energy of a free particle with mass $m_{\rm eff} = \hbar^2 /
2ta^2$.  The total energy (classically) would therefore be given by
\be
\epsilon = -2t + {p^2}/{2m_{\rm eff}} + V_{\rm trap}(x);
\ee
using this relation as the orbit equation, and applying the WKB approximation as
in the harmonic case \cite{Hooley-Quintanilla-04},
one can show that the DOS near the bottom of the band has the form
\be
\rho(\epsilon) \sim \left( \epsilon + 2t \right)^{(1/n)-(1/2)}. \label{lowdos}
\ee
This produces the correct square-root singularity in the $n \to \infty$ limit, 
and the previously calculated constant behaviour in the
harmonic trapping case.  Notice that, according to 
(\ref{lowdos}), a singularity exists at the bottom of the band for 
any $n > 2$.

To explore what happens at the top of the band ($\epsilon \sim 2t$), we evaluate
the DOS in the LDA approximation (\ref{lda}) numerically.  The
results for a few different values of $n$ are shown in Fig.~\ref{differentn}.

In summary, we have presented an examination of the DOS of a
single atom in a strong optical lattice plus a shallow power-law confining
potential.  Varying the power law permits one
to scale from the case of harmonic trapping to the hard-wall case relevant to
solid-state physics, and we call this variation finite-curvature scaling.  The
next step, for fairly weakly interacting atoms, would be to introduce the
atom-atom interactions perturbatively.  In the fermionic case, one can see that
something special will happen (enhanced susceptibilities, or perhaps even
ordering) when the chemical potential lies near the singularity at $\epsilon =
2t$.  Analysis of this case is the subject of current research.

\section*{Acknowledgement}

We thank M.~Foulkes, N.~I.~Gidopoulos and A.~J.~Schofield for useful
discussions. JQ thanks the University of Birmingham for hospitality
while some of this work was carried out. We acknowledge financial
support from EPSRC (CH), the Leverhulme Trust and CCLRC (JQ).

\end{document}